\documentclass[intlimits,twoside,a4paper]{article}

\usepackage[cp1251]{inputenc}
\usepackage[eqsecnum]{cmpj3}

\usepackage{graphicx}
\usepackage{float}
\usepackage[caption = false]{subfig}

\issue{2022}{25}{4}{43708}
\doinumber{10.5488/CMP.25.43708}

\title[Quantum geometrical effects in KDP crystals]%
{Path integral Monte Carlo simulations of the geometrical effects in KDP crystals%
}
\author[F. Torresi, J. Lasave, S. Koval]{F. Torresi\orcid{0000-0002-4999-0530}, J. Lasave\orcid{0000-0003-1393-1114},
	 S. Koval\orcid{0000-0002-6925-4266}\thanks{Corresponding author: \email{koval@ifir-conicet.gov.ar}.}}
\address{
	 Instituto de F\'{\i}sica Rosario, Universidad Nacional de Rosario and CONICET, 27 de Febrero 210 Bis, 2000 Rosario, Argentina
}


\date{Received July 10, 2022}

\begin{document}

\maketitle

\begin{abstract}
Path integral Monte Carlo (PIMC) simulations with very simple models were used in order to unveil the physics behind the isotope effects in H-bonded ferroelectrics.
First, we studied geometrical effects in the H-bonds caused by deuteration with a general three-site model based on 
a back-to-back double Morse potential plus a Morse potential between oxygens, fitted to explain different general features for a wide set of H-bonded compounds.
Our model results show the Ubbelohde or geometrical effect (GE), i.e., the expansion of the H-bond with deute\-ration, in agreement to what is observed in H-bonded ferroelectrics with short H-bonds.
Moreover, adjusting the potential parameters to ab initio results, we have developed a 1D model which considers the bilinear proton-proton interaction 
in mean-field to study nuclear quantum effects that give rise to the GE in KDP crystals. PIMC simulations 
reveal that protons tunnel more efficiently than deuterons along the 1D chain, giving rise to a strong attraction center that
pulls the oxygens together. This mechanism, which is based on the correlation between tunneling and geometrial modifications of the H-bonds,
leads to a strong GE in the ordered phase of the chain at low temperature which is in good agreement with the experimental data.

\printkeywords

\end{abstract}

\section{Introduction}

KH$_2$PO$_4$ or KDP is the prototype of a wide family of
H-bonded ferroelectric compounds which has
extensive applications as a key component in optoelectronic devices~\cite{Lin77}.
Besides the technological interest, KDP has also attracted much attention
due to its rich, complex and intriguing phenomenology, e.g., the huge isotope effect that displays associated to its 
ferroelectric-paraelectric (FE-PE) phase transition.
With deuteration, the critical temperature $T_c$
changes from $\approx 122$~K to $\approx 210$~K.
The saturated polarization $P_s$ at low $T$ also shows a large isotope effect, increasing from $\approx 5.0$~\textmu{}C/cm$^2$ for KDP 
to $\approx 6.2$~\textmu{}C/cm$^2$ for a sample with 98\% of 
deuteration~\cite{Sam73}.

The origin of these strong isotope effects is still controversial.
The first explanation of the large increase
of $T_c$ upon deuteration was given by the quantum tunneling model~\cite{Bli60},
which focuses purely on mass-dependent effects.
However, increasing experimental evidence since
the late nineteen eighties showed that the large isotope effect
is mainly driven by geometrical modifications
of the H bonds~\cite{McM90,Nel91} (Ubbelohde effect~\cite{Rob39}).
The recent observation of tunneling in the PE phase
of KDP by neutron Compton scattering experiments
added even more controversy to the problem~\cite{Rei02},
although in deuterated KDP (DKDP), tunneling could
not be detected~\cite{Rei08}.

Ab initio calculations have recently shown that
tunneling and geometric effects are complementary aspects of the same phenomenon\cite{Kov02,Kov05}. With a simple selfconsistent
model based on ab initio results, it is
demonstrated that the wave function solution of
the nonlinear Schr\"odinger equation for deuteron/proton clusters evolves from a double peak to a broad single peak located
at the center of the H-bonds as the cluster mass diminishes.
This is explained by a strong nonlinear feedback
between proton delocalization (tunneling) and the effective proton potential
barrier in the H-bonds, which changes concomitantly with the H-bond geometry. 
It is concluded that such a large mass dependence can explain the
large isotope effect found in KDP, via an amplified and selfconsistent geometric modification of the H bond
in agreement with experiments.
On the other hand, these results are in striking contrast with the very weak dependence obtained at fixed potential and geometry.
Thus, the proton tunneling subunit
and the host lattice are strongly coupled and 
the host-and-tunneling system is not separable.

Many models were successfully developed in the past to shed light into the general phenomeno\-logy of H-bonded ferroelectric materials 
\cite{Bli66,Koj88,Sug91,Shc99,Mer02,Shc06,Lev09,Las09}.
In this paper, we address with very simple models the problem of geometrical effects in KDP crystals by performing path integral Monte Carlo (PIMC) simulations. 
First, we develop a three-site model for the H-bond to study local quantum geometric effects. This simple model already serves us to gain
knowledge about the interplay between proton tunneling and H-bond geometric modifications such as the O--O distance variation.
After this first insight, we develop a~1D chain model of concatenated H-bonds to study in the ordered phase the geometrical
effects caused by deuteration. The model parameters are fitted using recent ab initio results~\cite{Men18}. 
We demonstrate that this simple linear model can account for the geometrical effects observed in real H-bonded
ferroelectrics, which are at the root of the giant isotope effect in the critical temperature observed in the FE phase transitions of these materials. 
The paper is organized as follows: in the next section we explain the models used and describe details of the PIMC calculations. Section 3
describes and discusses the results obtained for the three-site model and for the linear chain. Finally, we elaborate 
a summary and our conclussions in section 4.

\section{Models and calculation details}

\subsection{Three-site model}

\begin{figure}[htb]
\centerline{{\includegraphics[width = 2.5in]{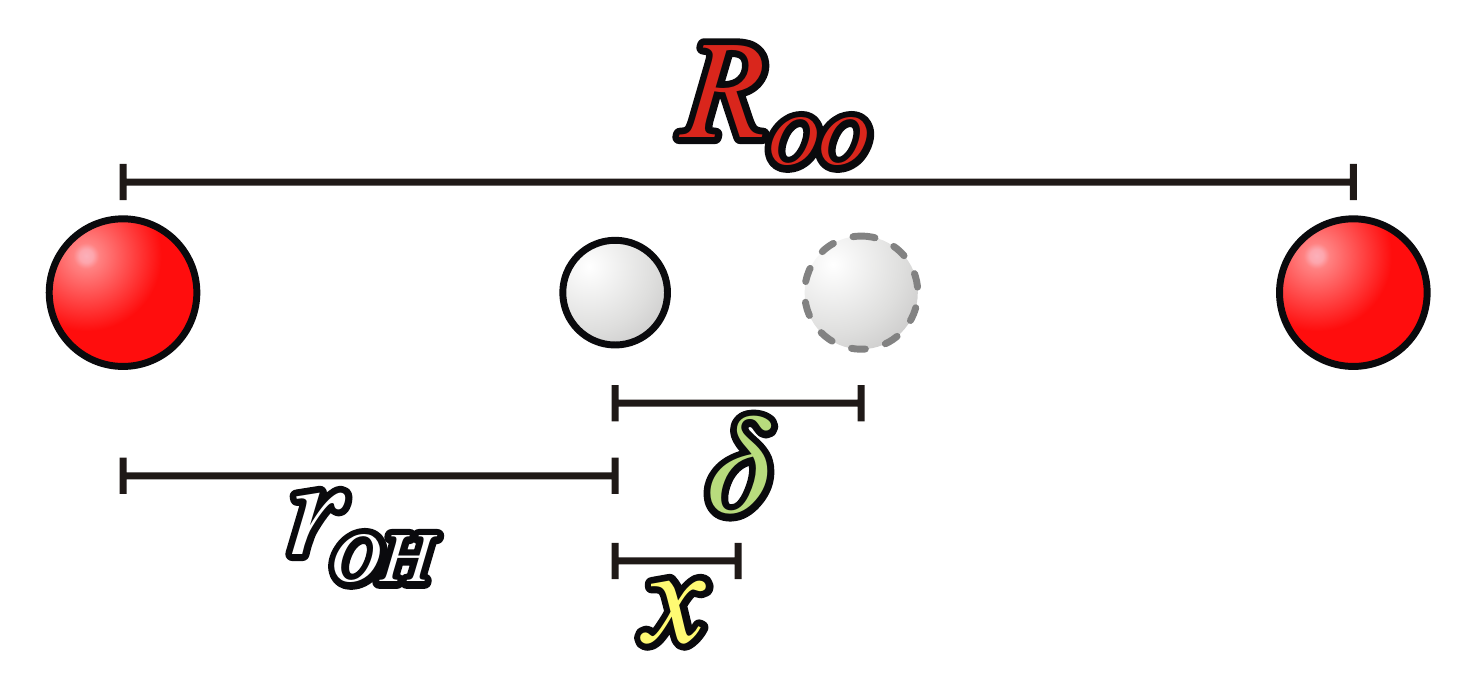}}}
	
	\caption{(Colour online) H-bond parameters in the three-site model. $R \equiv R_{\rm {OO}}$ is the distance between oxygen nuclei.
		$r_{\rm {OH}}$ is the proton-oxygen distance. The variable $\delta=R_{\rm {OO}}-2r_{\rm {OH}}$ 
		is defined as the distance between the two possible equilibrium positions of the proton. Then, $x=R_{\rm {OO}}/2-r_{\rm {OH}}$  
        is the proton coordinate relative to the H-bond center. This parameter definition is also used in the linear chain model.}
	\label{H-bond}
\end{figure}

We developed a three-site (3S) model which represents a single O--H--O cluster embedded in the H-bonded ferroelectric as it is sketched in figure~\ref{H-bond}.
With the aim to model linear H-bonds, a Double Morse (or back-to-back) potential (see e.g.,~\cite{Mat82,Mat82_4,Tan90,Yan93,Sci08}) is usually used, which is essentially the superposition of two Morse 
potentials representing what the
proton feels while interacting with both oxygens:
\begin{align}
V_{\rm {OH}} \left(x,\:R\right) & = V_{\scriptscriptstyle M} \left(x+\frac{R}{2}\right) +V_{\scriptscriptstyle M} \left(\frac{R}{2}-x \right) 
\nonumber\\
& =  D\left\{1 - {\rm {exp}}\left[{-a \left(\frac{R}{2} + x - r_{0} \right)}\right] \right\}^2
+ D\left\{1 - {\rm {exp}}\left[{-a \left( \frac{R}{2} -x - r_{0} \right)}\right] \right\}^2 - 2D,
\label{voh}
\end{align}
\noindent where $R$ is the O--O distance, and $x$ represents the H position relative to the H-bridge center (see figure~\ref{H-bond}).
If we assume that $R$ is fixed, there is a critical value $R_c=2(a^{-1}\ln 2 + r_{0})$ such that for $R < R_c$
the potential profile is a single well with a minimum at $x=0$. On the contrary, for $R > R_c$ we have a symmetric double-well potential,
with a local maximum at $x=0$ and minima at $x=\pm a^{-1} \cosh^{-1}\lbrace 1/2 \, {\rm {exp}}[a(R/2-r_{0})]\rbrace$.
Notice that the energy barrier for the proton jump from one side to the other of the H-bond diminishes concomitantly with the
O--O distance $R$, vanishing for $R < R_c$. Actually, we are interested in the proton/deuteron tunneling regime, thus we would 
need that the equilibrium distance $R$ remains in the region where the proton barrier exists, that is $R > R_c$. However,
simulations at low temperature with the potential described in equation~\ref{voh}, relaxing both variables $x$ and $R$, 
yield to a collapse of the potential barrier and the equilibrium energy profile displays one minimum only. 
Therefore, it is mandatory to introduce a new interaction which preserves the system from the O--O distance collapse.
This O--O potential will represent the interaction between both oxygens and the lattice.
The following Morse potential between oxygens is chosen~\cite{Men18}:
\begin{align}
V_{\rm {OO}} \left(R\right) =
D_{\rm {OO}}\;\left[1 - \re^{-a_{\rm {OO}} \left(R - R_{0} \right)} \right]^2 - D_{\rm {OO}}.
\label{voo}
\end{align}
We adopted a Morse potential to describe the O--O interaction with the lattice because
this kind of anharmonic potential enables the system to explore with sufficient probability
O--O distances larger than~$R_{0}$, in such a way that the collapse tendency to a single well
is drastically diminished. This is in contrast to the case of a harmonic potential
for the O--O interaction, where in this case the O--O collapse is inevitable.
The complete potential for the 3S model is as follows:
\begin{align}
V_{\scriptscriptstyle 3S} \left(x,R\right) = V_{\rm {OH}} \left(x,R\right) + V_{\rm {OO}} \left(R\right) &  =
 D\left\{1 - \re^{-a \left[({R}/{2}) + x - r_{0} \right]} \right\}^2 + D\left\{1 - \re^{-a \left[({R}/{2}) -x - r_{0} \right]} \right\}^2
 \nonumber\\
& - 2D + D_{\rm {OO}} \left\{1 - \re^{-a_{\rm {OO}} \left(R - R_{0} \right)} \right\}^2 - D_{\rm {OO}}.
\label{v3s}
\end{align}
The correlation between the H displacement $x$ and 
the O--O distance $R$ observed in experiments and ab initio calculations is reflected by the anharmonic potential of equation~(\ref{v3s}):
when the H approaches one of
the O's in the covalent bond O--H (increasing $x$), the hydrogen-bond with the other O weakens and the~O--O 
distance ($R$) increases. Moreover, $R$ diminishes with decreasing $x$, which is the inverse situation.
This correlation is precisely the important ingredient necessary for the existence
of the Ubbelohde or the geometrical effect observed in compounds with strong H-bonds.

\subsection{1D model of concatenated H-bonds}

\begin{figure}[htb]
	\vskip -0.6 cm
	\centerline{{\includegraphics[width = 5.5in]{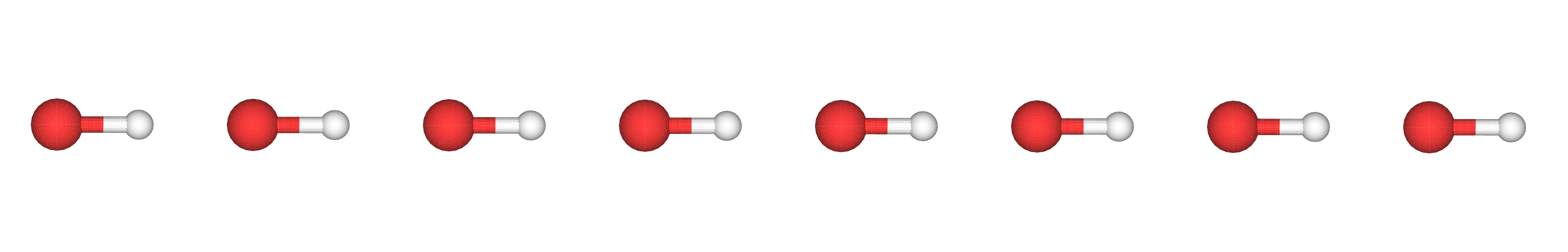}}}
	\caption{(Colour online) Schematic representation of the 1D chain model in the ordered phase. Each unit cell is formed with one oxygen (red sphere) and one hydrogen (white sphere).
		Our model consists of a supercell subjected to periodic boundary conditions containing 200 unit cells (for better visualization only 8 unit cells are shown).}
	\label{H-bonds-chain}
\end{figure}

Going a step beyond the simple three-site model, we have developed a one dimensional chain model of concatenated H-bonds to study the GE in a more realistic way in the ordered phase. This 1D linear model consists of a chain ...O--H...O--H...O--H...O--H...,
which is built as a supercell containing $N=200$ unit cells of linear dimension $R$, the O--O distance, as shown schematically
in figure~\ref{H-bonds-chain}.
There are two atoms, one oxygen and one hydrogen
in each unit cell (O--H...). The supercell of dimension $L=200R$ is subjected to periodic boundary conditions. In the simulation, $L$ is allowed to relax at zero stress,
as well as each coordinate $x_i$ and $R_i$ of each unit cell $i$. For instance, this chain represents a model approximation to the 1D H-bonded
ferroelectric CsH$_2$PO$_4$ (CDP) if the model chain oxygen is interpreted as a PO$_4$ unit plus an ordered hydrogen covalently bonded to the phosphate at any temperature, 
and the model hydrogen is the one that is
disordered at high temperature in CDP~\cite{Las16}. Then, the global motion of hydrogens in our linear model in the ordered phase, 
from one minimum to the other along the H-bonds of the chain,
could be related to the FE mode that accounts for the spontaneous polarization arising along the $b$ direction at low $T$ in CDP~\cite{Las16}. 
Alternatively, the chain model may also represent an approximation to the study of the GE in KH$_2$PO$_4$ (KDP) if the model effective oxygen now represents
a KDP cluster of two phosphate units including seven protons moving coordinately as a local FE mode~\cite{Kov02,Kov05}. In all these cases, we
must adopt a convenient effective mass for the effective model hydrogen/deuteron considering that the real displacements of H(D) are accompanied with 
the heavier atom motions~\cite{Kov02,Kov05,Men18}.

The total potential energy for the linear chain (1D) model is defined as:
\begin{align}
V_{\scriptscriptstyle 1D} \left(R\right) = \sum_i V_{\scriptscriptstyle 3s} \left(x_i,\:R_i\right) - \frac{1}{2} \sum_{\langle ij \rangle} J x_i x_j,
\label{1Dm}
\end{align}
\noindent where $ V_{\scriptscriptstyle 3s} $ is the unit cell local potential defined exactly in the same way for the 3S model,
as is shown in equation~(\ref{v3s}), and the last term is the short-range interaction energy between protons/deuterons stemming from the ice rules restrictions, 
i.e., in this 1D model, only one proton is attached to each oxygen. The last sum in equation~\ref{1Dm} is restricted to nearest neighbours for each index $\langle ij \rangle$.
There is no long-range part in this model, which precludes a phase transition in one dimension.
However, the last bilinear term is treated in mean-field, which enables the system to have a second order phase transition at finite temperature~\cite{Koe76}.
Therefore, the 1D model total potential, is written in the following way~\cite{Tuc10}:
\begin{align}
V_{\scriptscriptstyle 1D} \left(R\right) = \sum_i V_{\scriptscriptstyle 3s} \left(x_i,\:R_i\right) -  J \langle x \rangle \sum_{i} x_i  + \frac{1}{2} N J {\langle x \rangle}^2,
\label{1Dmmf}
\end{align}
\noindent where $\langle x \rangle \equiv 1/N \sum_i x_i$ is the time and lattice average of the $x_i$ positions for each unit cell $i$ taken at each MC step in the simulation.

\subsection{Path integral Monte Carlo simulations}

In the PIMC simulations~\cite{Cha97}, the effective short-time propagator for two 
adjacent points in the discretized imaginary-time path describing each 
quantum particle was evaluated to
fourth-order accuracy with the Takahashi-Imada approximation~\cite{Tak84,Cha97,Weh98}. The 
effective action in this case allows us
to significantly reduce the Trotter number $M$ required for
convergence. In all the simulations performed we have used
$M = 128$ beads for the quantum polymer associated with each
atom in the O--H...O bonds, which yielded well-converged
results~\cite{Cha97,Las16,Men18}. Additionally, a normal-mode representation of
the quantum polymers was used in order to ensure ergodicity
in the MC sampling~\cite{Cha97,Weh98}. The PIMC simulations were
performed at low $T = 50$~K such that the quantum nuclear effects were predominant compared
to entropic contributions in the 3S model and also with the aim to obtain GE in the ordered phase for the 1D model
(the classical version of this model has a transition to a disordered paraelectric phase at~$\approx 350$~K).
The simulations for the 3S model consisted of $1 \times 10^6$ MC steps preceded by $5 \times 10^5$ steps of
thermalization. In the 1D chain model simulations, we took $3 \times 10^4$ steps of thermalization 
plus $1 \times 10^5$ MC steps for computing averages. In this case, each calculation performed
was an average of 20 runs with different random number generator seeds.

To characterize the degree of particle delocalization in the PIMC simulations, we studied the centroid
and radius of gyration (RG) distributions for the quantum polymers~\cite{Mor09}. The centroid
is defined as the center of mass (CM) of the polymer and represents the average position of the quantum particle.
The radius of gyration represents the variance of the quantum path and is a quantitative measure of how far away
are the beads or monomers from the polymer center, and therefore, provides a measure of the quantum 
delocalization of the particle~\cite{Mor09}.

\section{Results and discussion}

\subsection{Geometrical effect study using the three-site model }

The six potential parameters of equation~(\ref{v3s}) have been fitted in order to perform the GE study with the 3S model.
First, we fixed the values of  $a=2.89 \, \text{\AA}^{-1}$~\cite{Mat82,Mat82_4} and $D=3.12$~eV of the model parameters for the proton potential 
defined in equation~\ref{voh}, such that the stretching frequency for the O--H bond in the limit $R \rightarrow \infty$ coincides
with the experimental average value $\omega_{\infty} \approx 3750$~cm$^{-1}$~\cite{Mat82,Mat82_4,Mck14} for different H-bonded compounds.
There is a strong correlation between the OH and OO distances for the family of H-bonded compounds. The equilibrium distance $r_{\rm {OH}}$ 
diminishes systematically with increasing $R$ for $R > R_c$~\cite{Ich78,Jos82}, reaching a saturated value around $r_{\rm {OH}}^{\infty} \approx 0.95 \, \text{\AA}$
for very large $R$. Therefore, we took the parameter value $r_{0}=0.93 \, \text{\AA}$ so that the values $x$ that minimize 
$V_{\rm {OH}} \left(x,R\right)$ in equation~(\ref{voh}) for different values of $R$ give a curve $r_{{\rm {OH}}}^{{\rm min}}=R_{{\rm {OO}}}/2-x^{{\rm min}}$ as a function of $R$ that is a lower bound for the set of experimental points spread in the OH--OO correlation~\cite{Ich78,Jos82,Mat82,Mat82_4}.
With this choice, when the nuclear quantum effects are included in the PIMC calculations, we observe a very good agreement with the experimental
correlation curve using the model of equation~(\ref{voh}) with the OO distance $R$ fixed~\cite{Tor22}.

On the other hand, the parameter values for the OO interaction $V_{\rm {OO}} \left(R\right)$ [see equation~(\ref{voo})],
were initially taken from reference~\cite{Yan93}. They were further adjusted, especially the value of $D_{\rm {OO}}$, due to the 
important correlation between $r_{\rm {OH}}$ and $R_{\rm {OO}}$, such that the classic potential profile has the minimum at $R^{{\rm cl}}_{\rm {OO}} \approx 2.55\, \text{\AA}$.
We considered this condition because the most important geometrical effects are observed in H-bonded crystals with strong H-bonds which
have distances $R$ in a range between 2.5 and~2.6 $\text{\AA}$~\cite{Ich00}, with $R^{{\rm cl}}_{\rm {OO}}$ lying precisely in the middle of that window.
The final parameter values for the 3S model are shown in table \ref{6params}.

\begin{table}[!h]
	\begin{center} 
		
		\caption{Potential parameters used in the 3S model.}
		\label{6params}
		\vspace{2ex}
		\begin{tabular}{|c|c|c||c|c|c|}
				\hline
				$D$ [eV] & $a$ $[${ \AA{}}$^{-1}]$ & $r_{0}$ $[${ \AA{}}$]$ &  $ D_{\rm {OO}}$ [eV] & $a_{\rm {OO}}$ $[${ \AA{}}$^{-1}]$ &	$R_{0}$ $[${ \AA{}}$]$            \\
				\hline
				$3.12$ & $2.89$ & $0.93$  &  $0.55$ & $2.28$ & $2.76$        \\
				\hline
			\end{tabular}
	\end{center}
\end{table}

\begin{figure*}[tbp]
	\centerline{\subfloat[Proton]{\includegraphics[scale=0.4]{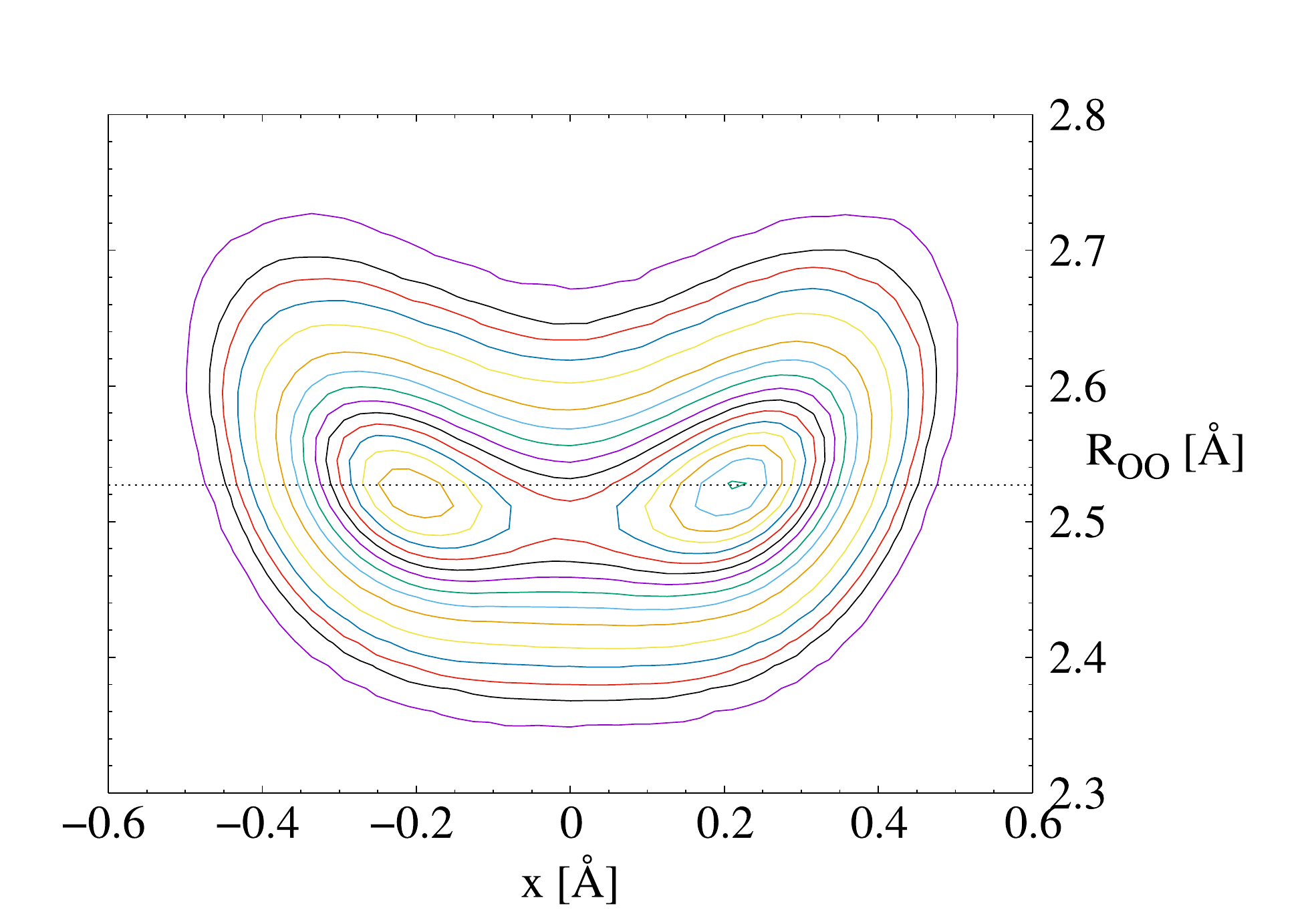}} 
\hskip -1.0cm
	\subfloat[Deuteron]{\includegraphics[scale=0.4]{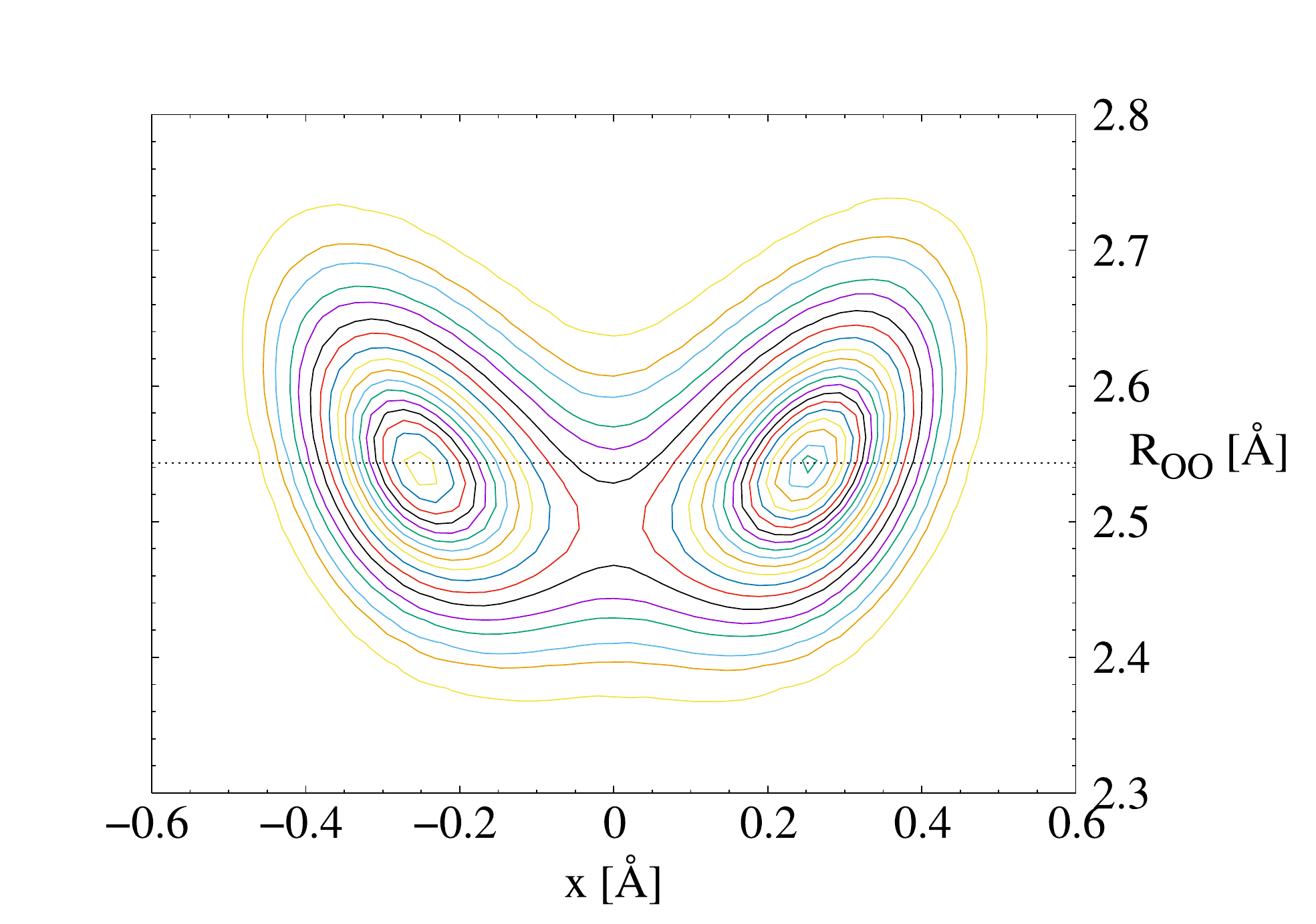}}}
	
	\caption{(Colour online) Proton/Deuteron probability distribution contours for the three-site PIMC simulations at $T=50$~K.}
	\label{pddist_3s}
\end{figure*}

We have verified that the 3S-model PIMC simulations performed at $T=50$~K with $M=128$ beads for the quantum polymer representing each
atom yielded probability distributions for the H-bond parameters~($x$ and $R$) and energies well converged.
The low temperature of 50~K for the simulation was chosen because we are interested in the nuclear quantum
effects for the H-bonds and the geometrical changes with deuteration without most of the influence of entropic contributions in the particle dynamics.
The 3S model results for the probability density contours to find the system in a given $(x,R)$ configuration
are shown in figure~\ref{pddist_3s} for the proton and deuteron cases.
The curves are qualitatively different but both cases are found to have symmetric distributions 
around $x=0$ in the $x$ coordinate with two prominent peaks with maximum probability, which are clearly shifted in the deuterated case.
The OO distance for the peak positions are in each case: $R_{\rm {OO}}^{{\rm peak}}(H)= 2.527\, \text{\AA}$ and
$R_{\rm {OO}}^{{\rm peak}}(D)= 2.543\, \text{\AA}$, which represents a distance enlargement for the OO bond of
$\Delta R_{\rm {OO}}=0.016 \, \text{\AA}$, evidencing the geometrical or Ubbelohde effect of the H-bond
expansion with deuteration. Moreover, the corresponding average values also increase with deuteration:
$\langle R_{\rm {OO}}(H) \rangle =2.525 \, \text{\AA}$ and $\langle R_{\rm {OO}}(D)\rangle =2.540 \, \text{\AA}$.

The PIMC simulations also show a change in the variable $\delta$ with deuteration
for the peaks observed in figure~\ref{pddist_3s}. The variation is: $\Delta \delta= \delta_{D} - \delta_{H}= 0.079 \, \text{\AA}$,
where $\delta_{H}=0.417 \, \text{\AA}$ and $\delta_{D}=0.496 \, \text{\AA}$. 
This is also reflected in a shrinking of the O--H bonds: $\Delta r=r_{\rm {OH}}-r_{\rm {OD}}=0.032  \, \text{\AA}$.
The overall changes in the variables $\delta$ and $R$ with deuteration in the simulations are in agreement with
what is observed in the experimental data for different H-bonded compounds with strong H-bonds~\cite{Sok88,Ich00}. Thus,
our simple 3S model satisfactorily reproduces the isotopic geometrical effects for these systems.

It is worth to notice that if the OO distance is not allowed to relax, then the GE is smaller. For instance, we have fixed the value 
$R_{\rm {OO}}=2.527 \, \text{\AA}$, which corresponds to the peak in the probability distribution for the protonic system (see figure~\ref{pddist_3s}),
and the simulations gave a change with deuteration in the OH bond of only $\Delta r = 0.021\,  \text{\AA}$.
Comparing this result with that considering the oxygen dynamics ($\Delta r=r_{\rm {OH}}-r_{\rm {OD}}=0.032  \, \text{\AA}$), we observe 
an increment of $\approx$ $50 \%$ in the isotopic geometrical effect in the case where the oxygens are allowed to relax. 
This can be understood in the following way: first, when the oxygens are fixed, protons, being more
delocalized than deuterons, have more probability to stay closer to the middle of the O--O bond.
Second, when the oxygen dynamics is included, the protons act as a strong attraction center that pulls
the two bridge oxygens together, more effectively than deuterons which are more localized near the oxygen.
This proton-mediated O--O contraction lowers the potential barrier, which delocalizes even more the proton, and so on,
giving rise to a nonlinear selfconsistent mechanism~\cite{Kov02,Kov05}. For the deuteron, being less delocalized than the proton,
the selfconsistent effect is weaker. This mechanism leads to an isotopic geometrical effect 
which is stronger than that generated by the proton/deuteron quantum delocalization at fixed potential (fixed oxygens)~\cite{Kov02,Kov05}.

\begin{figure}[!tbp]
	\subfloat[Proton]{\includegraphics[width = 3in]{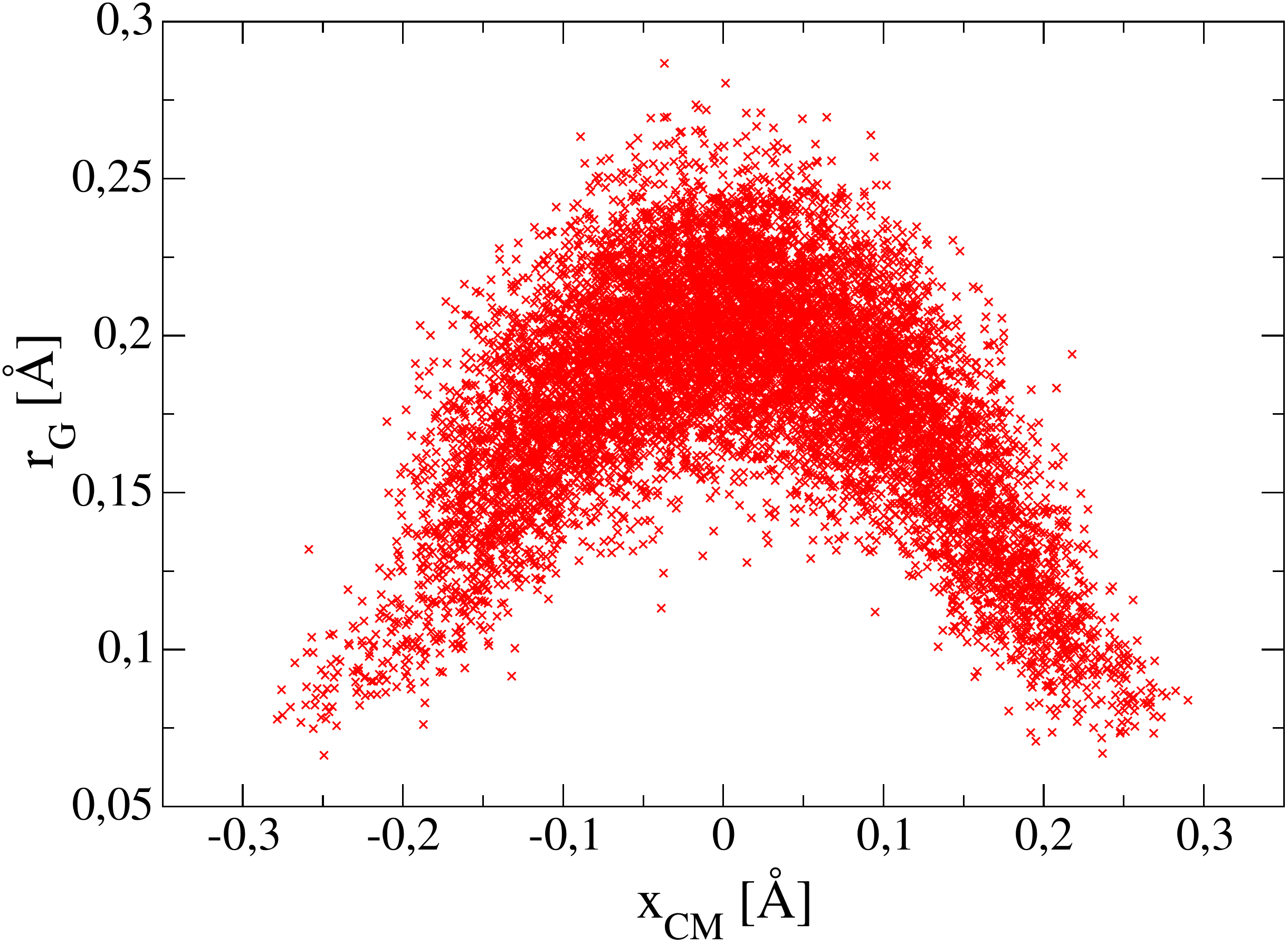}}
	\subfloat[Deuteron]{\includegraphics[width = 3in]{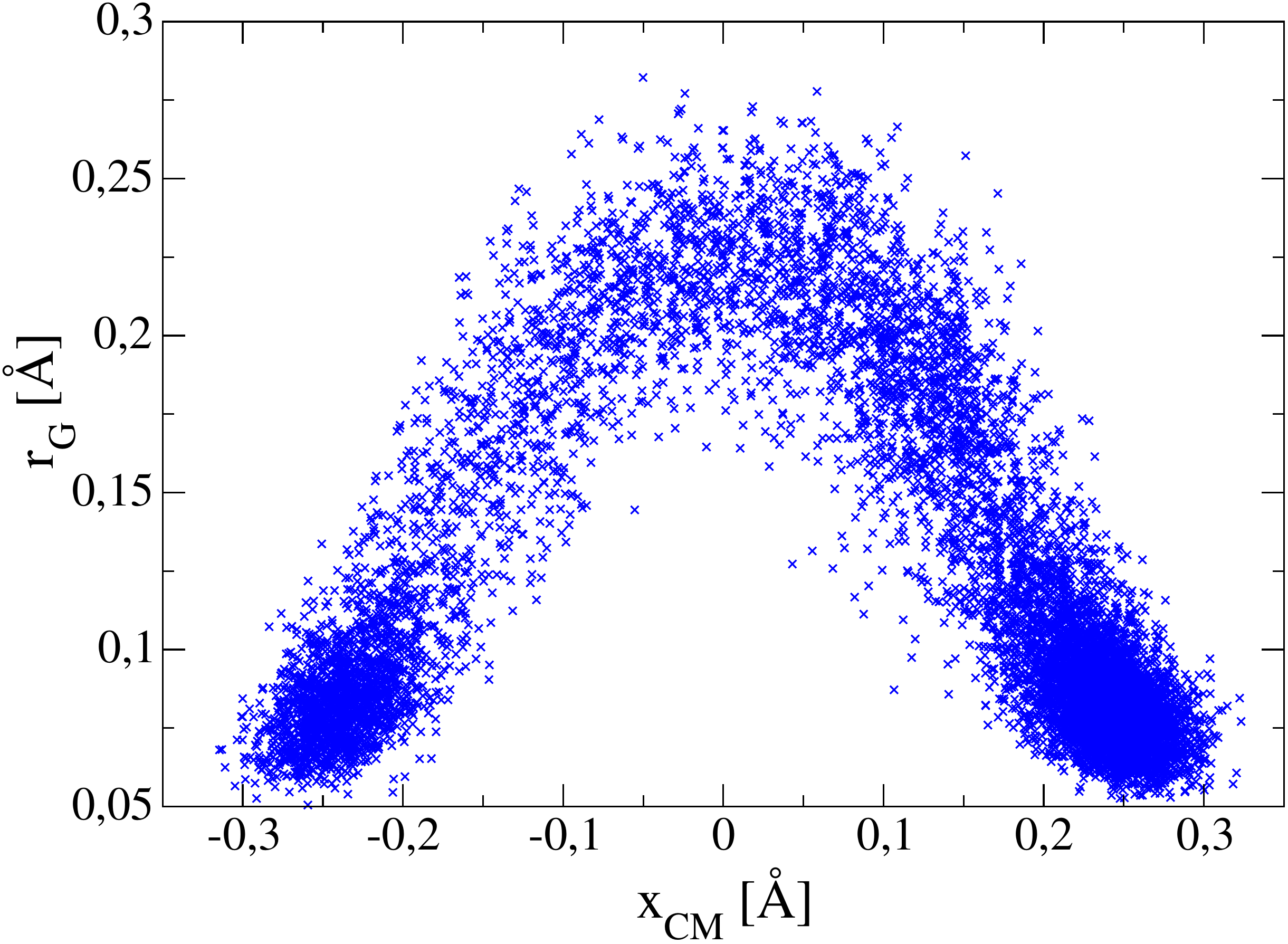}} 
	
	\caption{(Colour online) Distribution of the radius of gyration $r_G$ vs. centroid coordinate $x_{CM}$ for the three-site simulations at $T=50$~K.}
	\label{gyr_xm_3s}
\end{figure}

To further illustrate the microscopic mechanism that rules the GE, we have analyzed the 
behavior of the quantum polymers for the proton/deuteron in the simulation via
an analysis of the center of mass of the quantum polymer or centroid position $x_{CM}$ and the 
radius of gyration $r_G$ representing a measure of the quantum delocalization of the particle
(i.e., the extension of the quantum polymer)~\cite{Mor09}.
We plot in figure~\ref{gyr_xm_3s} the instantaneous values of $r_G$ as a function of
the proton/deuteron centroids $x_{CM}$, taken every~100~MC steps in the PIMC simulation.
As can be seen in the figure, the density of points reveals that 
the deuteron prefers to be localized at both sides and far from the bond middle 
with small values of~$r_G$, indicating a more classical behavior in these cases.
When the deuteron centroid takes the values of~$x_{CM}$ closer to 0 (the bond middle),
it is observed an increase of $r_G$ indicating that the quantum polymer is delocalized
and is spread through both sides of the potential barrier, signaling the presence of tunneling
in this case. Notice that the largest values of $r_G$ are found at $x_{CM} \approx 0$ where
delocalization is maximum. On the other hand, in the proton case, tunneling is much more
frequent because the region with larger density of points appears near $x_{CM} \approx 0$ with
large values of $r_G$, as shown in figure~\ref{gyr_xm_3s}.
This is precisely an important ingredient for the GE: the proton spends
much more time delocalized with the quantum polymer center of mass near the middle of the O--O bond, which finally produces
a strong contraction of the O--O distance. On the contrary, the deuteron is much more
localized at both sides and far from the bond middle which leads to a weakening of the O--O bond and to an
increase of the O--O distance. This yields the isotopic geometrical effect, which is observed
in the calculated probability distribution of figure~\ref{pddist_3s}.

\subsection{Isotope effects obtained with the 1D model simulations}

The previous analysis of the 3S model results, which 
has clearly shown the isotopic GE, was carried out
based on the parametrization of the model which reproduces the universal OH--OO correlation observed for a 
family of diverse H-bonded compounds. In this sense, this model is quite simple and general, accounting for
the geometrical effects with deuteration of a set of H-bonded ferroelectrics with strong H-bonds. 
Now, we focus on the development of a 1D chain model, described in section 2.2 [see equation~(\ref{1Dmmf})], 
which was specifically designed to explain the isotope effects in the phase transition of KDP and was fitted to ab initio results~\cite{Men18}. 
This more realistic 1D model has, in the classical nuclei version, a ferroelectric-paraelectric transition
at $T \approx 350$~K~\cite{Tor22}. In this paper, we have used it in the ordered phase of KDP at $T=50$~K to analyze the isotopic GE  
which is at the root of the microscopic mechanism that leads to the giant isotope effect in the critical temperature.

We start from equation~\ref{1Dmmf} for the 1D model, which has seven parameters to be adjusted for the KDP case.
The six model parameters of the local proton potential $V_{\scriptscriptstyle 3S}$ for each unit cell in the chain, 
which is just the same that was used in the 3S model (see equation~\ref{v3s}), 
have been adjusted to reproduce six magnitudes obtained from ab initio calculations for KDP. 
These magnitudes were the global energy barrier between the PE and FE states, 
the O--O and $\delta$ distances in the FE phase, the O--O distance in the PE phase, the ab initio vibrational frequency of the
PO$_4$ rotation mode, which is equivalent to the stretching mode in the 3S model, and
the energy barrier between the
energy minimum and the transition state in the FE
phase keeping the O--O distance fixed (see reference~\cite{Men18}). 
We adopted the model fit to the ab initio calculations that
includes dispersion corrections at the vdW-DF level, which
exhibit, compared to other methods, the best agreement with the experimental geometry for both
KDP and deuterated KDP (DKDP)~\cite{Men18}.

Finally, we have fitted the remaining parameter $J$ that corresponds to the proton-proton
interaction term in equations~(\ref{1Dm}) and (\ref{1Dmmf}). To this end, $J$ was adjusted to 0.55~eV/$\text{\AA}^2$
so that the critical temperature $T_c$ for the FE-PE transition obtained by the 1D model simulation with classical nuclei
reaches the value of $\approx$ 350 K, similar to the value obtained by ab initio molecular dynamics calculations
with dispersion corrections at the vdW-DF level for DKDP~\cite{Men19}.

The final values for the parameters used in the 1D model are listed in table  \ref{7params_1D}.

\begin{table}[H]
	\begin{center} 
		\caption{Potential parameters used in the 1D model.}
		\label{7params_1D}
		\vspace{2ex}
		\begin{tabular}{|c|c|c|c|c|c|c|}
			\hline
			$D$ [eV] & $a$ $[${\AA{}}$^{-1}]$ & $r_{0}$ $[${\AA{}}$]$ &  $ D_{\rm {OO}}$ [eV] & $a_{\rm {OO}}$ $[${\AA{}}$^{-1}]$ &	$R_{0}$ $[${\AA{}}$]$    &      $J$  [ eV/$\text{\AA}^2]$     \\
			\hline
			$8.838$ & $3.027$ & $0.966$  &  $10.542$ & $0.831$ & $2.917$  &   0.55       \\
			\hline
		\end{tabular}
	\end{center}
\end{table}

The motion of the proton/deuteron is strongly correlated with that of the 
heavy ions, and its mass is dressed accordingly as discussed in reference~\cite{Kov05}.
Therefore, instead of using the bare proton (deuteron) masses $m_p$ ($2m_p$), we have used in the PIMC simulations the effective 
masses for H and D: $\mu_H=2.3m_p$ and $\mu_D=3m_p$, 
respectively, with $m_p$ the proton mass~\cite{Kov02,Kov05,Men18}.

\begin{figure}[!tbp]
	\centerline{\subfloat[Proton]{\includegraphics[scale = 0.4]{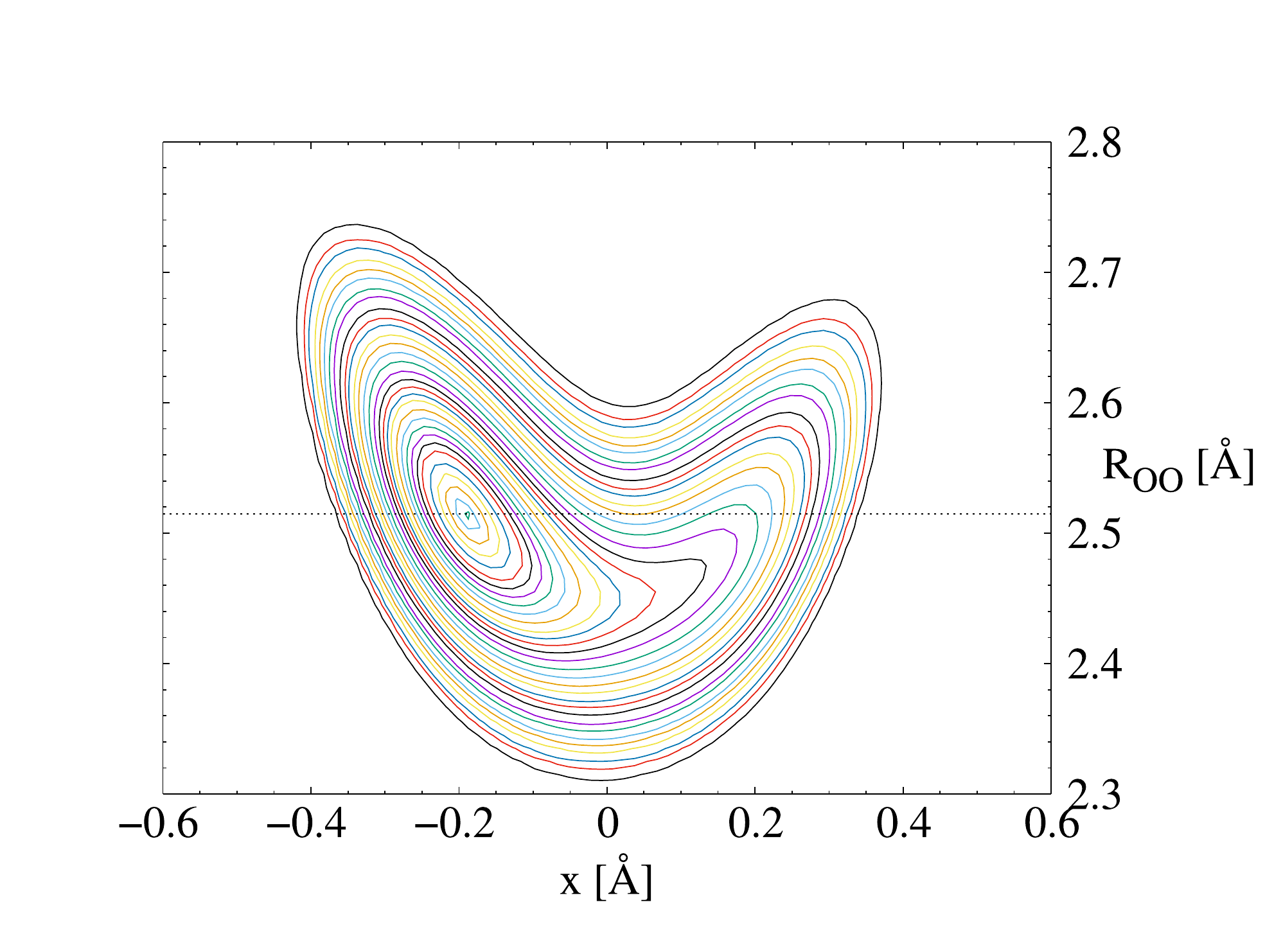}} 
	\hskip -0.8cm
	\subfloat[Deuteron]{\includegraphics[scale = 0.4]{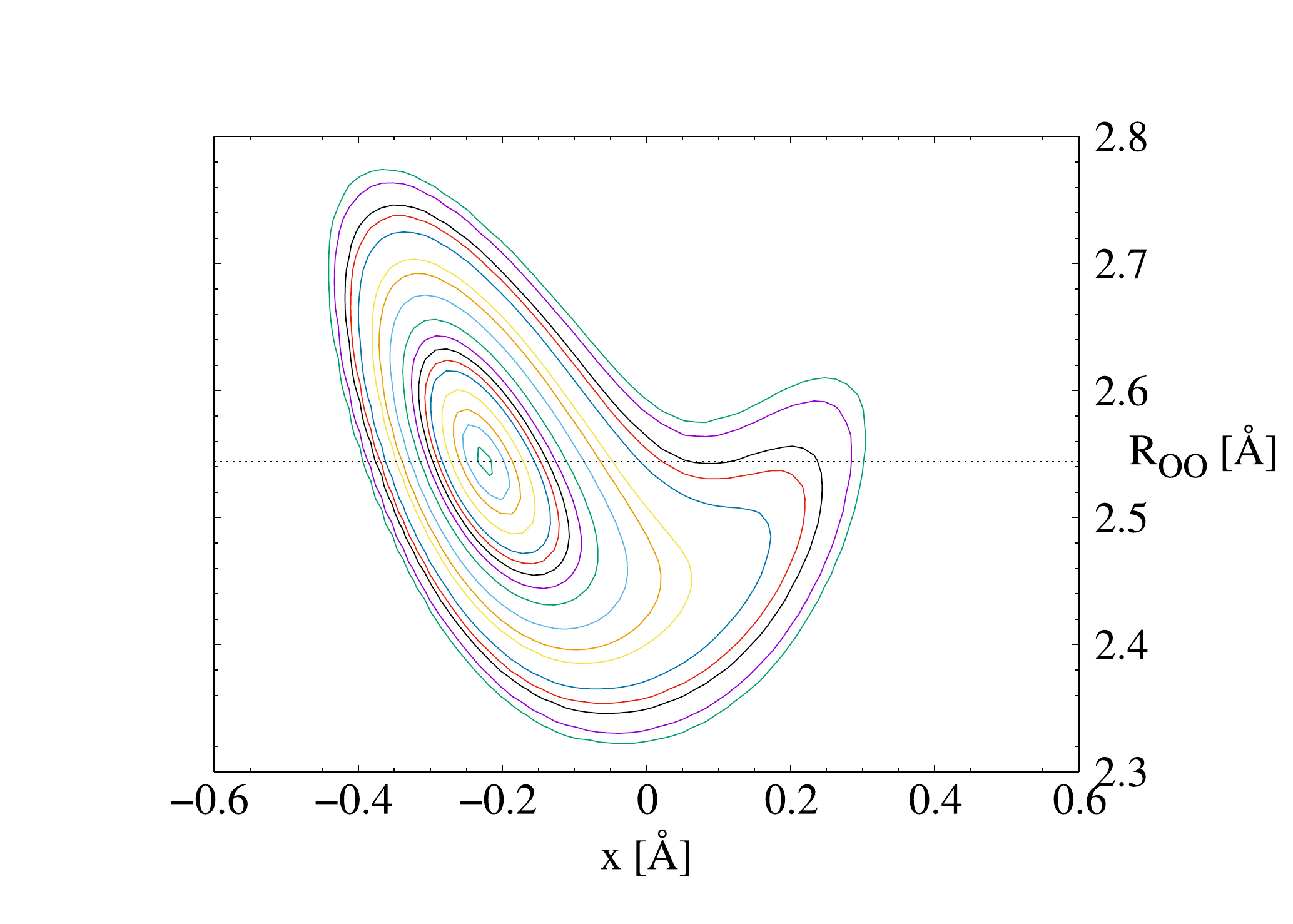}}}
	
	\caption{(Colour online) Proton/Deuteron probability distribution contours in the H-bonds for the linear-chain PIMC simulation at $T=50$~K.}
	\label{pddist_1d}
\end{figure}

We plot in figure~\ref{pddist_1d} the probability distribution contours for the PIMC simulation with the 1D model, obtained in the ordered phase at $T=50$~K. 
Due to the ordered phase, only one peak is observed in the proton and deuteron
distributions, which is in contrast to the symmetrical double peaks around~$x=0$ 
found in the 3S model distribution results (see figure~\ref{pddist_3s}). The calculated distribution for the
chain of protons in figure~\ref{pddist_1d} is asymmetric around the peak position due to the potential anharmonicity and quantum delocalization, which is in qualitative agreement with the experimental
diffraction pattern measured near $T_c$ in the FE phase of KDP~\cite{Nel85}. The asymmetry around the peak is less pronounced in the deuterated case
as shown in figure~\ref{pddist_1d}, because the deuteron is less delocalized than the proton.

The prominent single peak found in the distribution results for the 1D simulation is clearly shifted in the deuterated case towards larger $x$ and $R$, 
revealing the existence of the isotopic geometrical effect, i.e., the expansion of the H-bonds in the chain with deuteration. 
The O--O distance for the peak positions are in each case: $R_{\rm {OO}}^{{\rm peak}}(H)= 2.515\, \text{\AA}$ and
$R_{\rm {OO}}^{{\rm peak}}(D)= 2.542\, \text{\AA}$, which represents a distance enlargement for the O--O bond of
$\Delta R_{\rm {OO}} \equiv  R_{\rm {OO}}(D)  - R_{\rm {OO}}(H) =   0.027 \, \text{\AA}$.
The $x$ coordinate of the peak position also expands with deuteration, from $x_H^{{\rm peak}}= 0.188\, \text{\AA}$
to $x_D^{{\rm peak}}= 0.218\, \text{\AA}$, with a net increase of $\Delta x=0.030 \, \text{\AA}$ or similarly
$\Delta \delta  \equiv \delta_D - \delta_H = 0.060 \, \text{\AA}$.
These results are summarized in table~\ref{results_1d} and compared with the available experimental data for KDP and DKDP~\cite{Nel87}.
We observe a good agreement between theory and experiment, although the GE is a little bit underestimated, with difference values under 
deuteration $\approx 25$\% lower than the experimental data.

\begin{table*}[!h]
	\begin{center} 
	\caption{Nuclear quantum calculations of the H-bond geometries for KDP and DKDP using the 1D linear model.
	The results, which correspond to the peak positions of figure~\ref{pddist_1d}, are contrasted with the experimental data of reference~\cite{Nel87}.			Distances are in {\AA{}}.}
	\label{results_1d}
	\vspace{2ex}
	\scalebox{0.95}{\begin{tabular}{|c|c c | c c | c c|}
			\hline
			\hline
			{\bf  $\,$ PIMC } &  \multicolumn{2}{c|}{\bf KDP ($\mu_H=2.3 \, m_p$)} &  \multicolumn{2}{c|}{\bf DKDP ($\mu_D=3.0 \, m_p$)} 
			                                             &  $\Delta R_{\rm {OO}}$   &    $\Delta \delta $                    \\
			{\bf results}  & {$R_{\rm {OO}}$} & {$\quad \delta \quad$} & {$R_{\rm {OO}}$}   & {$\quad \delta \quad$} &   &    \\
			\hline

			\bf{ 1D model 	} &  	2.515	& 	0.376	 &  	2.542	  & 	0.436   &  0.027  &  0.060	\\
		
			\hline
			{   \bf{Expt.~\cite{Nel87} } }  & 	2.497	 &    0.385	 &   2.533	 & 	0.472  &   0.036     &    0.087  	\\
			
			\hline
			
			\hline
			\hline 
		\end{tabular}}

		 \end{center}
	\end{table*}
	
\begin{figure}[tbp]
	\subfloat[Proton]{\includegraphics[width = 3in]{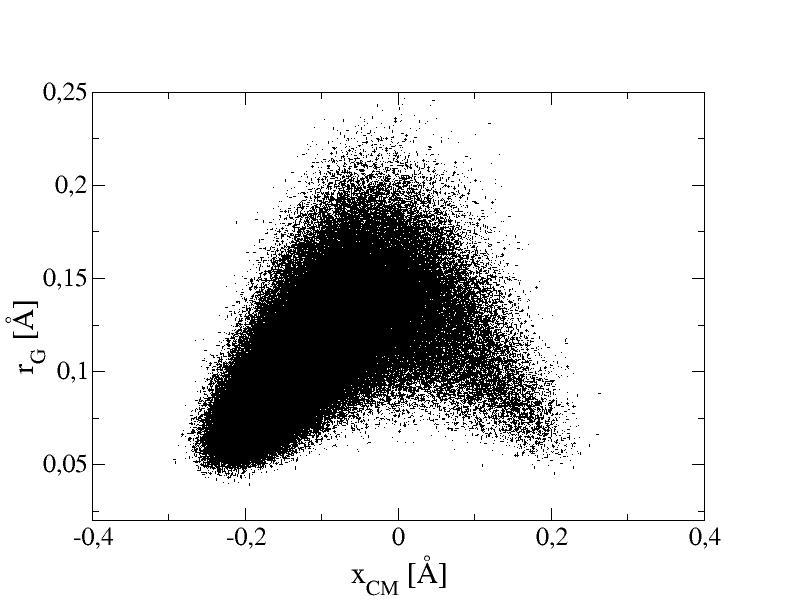}}
	\subfloat[Deuteron]{\includegraphics[width = 3in]{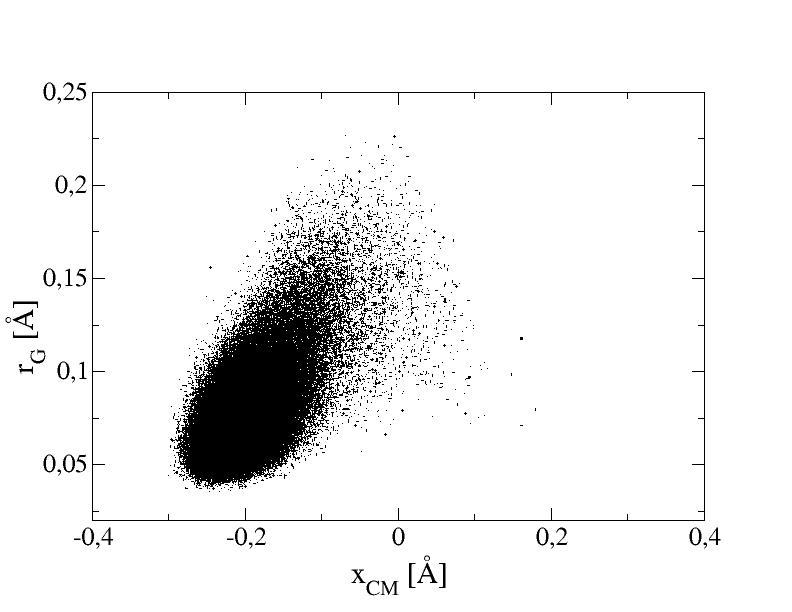}} 
	
	\caption{Distribution of the radius of gyration $r_G$ vs. centroid coordinate $x_{CM}$ of the quantum polymer representing
		the protons (a) and deuterons (b) relative to the center of the H-bonds, for the linear-chain PIMC simulation at $T=50$~K.}
	\label{gyr_xm_1d}
\end{figure}

To get a deeper insight into the microscopic mechanism of the geometrical effect in the linear chain model, we plot in figure~\ref{gyr_xm_1d} the distribution of the instantaneous radius of gyration $r_G$ as a function of the centroid positions $x_{CM}$ for all H-bonds in the chain,
where the points are taken every 100 MC steps along the PIMC simulation.
The region with largest density of points in figure~\ref{gyr_xm_1d} coincides with the position of the peaks in both proton
and deuteron cases (see figure~\ref{pddist_1d}). We again observe an asymmetric distribution 
centered in one of the sides of the H-bond consistent with the ($x,R$) distribution pattern of figure~\ref{pddist_1d}. 
The asymmetry observed in figure~\ref{gyr_xm_1d} is more pronounced in the proton case, indicating that protons jump more often than deuterons to
the other side of
the O--H--O bond.
The mechanism to pass through the potential barrier is to increase the radius of gyration near $x_{CM} \approx 0$ which means that
the particle tunnels through the barrier. This is helped by a strong contraction of the $R$ distance, which diminishes concomitantly with the potential barrier,
to a lower bound of $R_{{\rm min}} \approx 2.3\, \text{\AA}$ near $x=0$ as shown in figure~\ref{pddist_1d}. 
Thus, we conclude that tunneling is assisted by the $R$ distance modulation.
However, in this ordered phase at $T=50$~K, the proton spends more time in one of the sides of the O--H--O bond where the behavior is more
classic (low value of $r_G$). On the other hand, in the deuteron case, the particle remains localized practically all the time,
with a general classical behavior with low values of $r_G$. In other words, the tunneling for the deuteron is very scarce.
These results are consistent with the general assumption in the tunneling model: protons are capable of tunelling while deuterons are not~\cite{Bli60}. 
However, there is an essential difference: protons tunnel being assisted by the strong correlation with the O--O distance, which is
the behavior that originates the geometrical effect~\cite{Kov02,Kov05}. Therefore, the proton has a larger probability than the deuteron to spend more time tunneling
through the barrier near the middle of the O--H--O bond, and this generates a strong attraction center that pulls the two oxygens together, much more efficiently
than deuterons. This ``tunneling -- geometrical effect'' interrelation gives rise to the final geometrical effect observed in KDP crystals, that is, 
the H-bond expansion with deuteration,
which is crucial for the isotope effects in the FE-PE phase transitions~\cite{Kov02,Tor22}.

\section{Summary and conclusions}

We have carried out PIMC simulations with simple models to account for the geometrical effects~(GE) with deuteration in H-bonded ferroelectrics such as KDP crystals.
Firstly, we have developed a general three-site (3S) model consisting in a back-to-back double Morse potential for the O--H interaction and
a Morse potential which represents the interaction between the oxygens and the lattice. The model was fitted to reproduce general features
for a large set of different H-bonded compounds. The computed probability distribution contours in the ($R,x$) configuration space, with $R$ the O--O distance
and $x$ the proton/deuteron distance to the middle of the O--O bond, reveal a symmetric distribution around $x=0$ with two peaks on either side, for both proton and deuteron cases. The results show an increase with deuteration of $R$ and $x$ for the observed peaks, i.e., a GE,
which is in agreement with that observed in H-bonded compounds with strong H-bonds. Moreover, if the oxygens are not allowed to relax during the simulation,
the GE in the $x$ coordinate is much smaller, which means that there is a strong correlation between $R$ and $x$ that is important for the GE. During the PIMC simulations we have also plotted the instantaneous radius of gyration $r_G$ vs. the centroid position $x_{CM}$ of the quantum polymer 
representing the proton/deuteron. The results show that the proton tunnels more frequently than the deuteron (that is, it spends more time with the
center of mass near $x_{CM}=0$ with large values of $r_G$), while the deuteron is more localized in both sides and far from the O--H--O bond center,
with small values of $r_G$ (i.e., a more classsical behavior). These features yield a more effective contraction of the O--O bond in the proton case,
explaining the GE observed.

Secondly, we have developed a more realistic 1D model, with the same local potential for the H-bonds as that used in the 3S model, but
adding also a bilinear proton-proton interaction treated in mean-field. The parameters of the 1D model were fitted to ab initio results for KDP.
The bilinear interaction parameter of the model was adjusted such that the classical nuclei version of the model has a second order FE-PE phase 
transition at $T=350$~K in agreement with ab initio molecular dynamics simulations for DKDP. In this paper, by means of PIMC simulations of the 1D model, we have studied
the GE caused by deuteration in the ordered phase at $T=50$~K. The calculated probability distribution contours show only one peak in the ($R,x$) configuration
space for both proton/deuteron cases. The distribution is more asymmetric in the proton case due to the anharmonicity of the potential and 
the quantum delocalization. The distribution pattern is in qualitative agreement with the experimental distribution determined 
by high-resolution neutron diffraction studies~\cite{Nel85}.
The probability distribution contours show a peak which shifts substantially with deuteration. The changes in H-bond geomentry caused by the
GE observed in the~1D model
simulations are in good agreement with the corresponding experimental data.
The distribution of the radius of gyration vs. the quantum path centroids shows that the protons tunnel through
the potential barrier frequently while the deuterons are much more localized in one of the sides of the O--H--O bond and practically do not tunnel, 
in agreement with the
well-known tunneling model~\cite{Bli60}, and also with recent neutron Compton scattering experiments~\cite{Rei02,Rei08}.
We have shown that proton tunneling is assisted by a strong contraction of the O--O distance in the 1D model. Thus, there
is a strong correlation between instantaneous tunneling and geometrical effects of the H-bond that is much more
efficient in the proton case than in the deuterated system, which gives in average a strong GE for the whole simulation.
This mechanism is expected to be at the root of the huge isotope effect observed in H-bonded ferroelectrics of the KDP type~\cite{Kov02,Kov05}.

\section*{Acknowledgements}

We acknowledge support from Consejo Nacional de Investigaciones Cient\'{\i}ficas y T\'ecnicas (CONICET), Argentina.

\newpage
\ukrainianpart

\title{Метод інтегралів за траєкторіями у моделюванні Монте-Карло геометричних ефектів у кристалах  KDP}
\author{Ф. Торрезi, Х. Ласаве, С. Коваль}
\address{
	Інститут фізики Росаріо, Національний університет Росаріо та Національна рада з науково-технічних досліджень, вул. 27 лютого, 210 Bis, 2000 Росаріо, Аргентина
}

\makeukrtitle

\begin{abstract}
	\tolerance=3000%
	Метод інтегралів за траєкторіями у моделюванні Монте-Карло (ІТМК) для дуже простих моделей застосовано для з'ясування фізичних механізмів, що лежать в основі ізотопічного ефекту в сегнетоелектриках з водневими зв'язками. Зумовлені дейтеруванням геометричні ефекти у водневих зв'язках було досліджено за допомогою загальної тривузлової моделі, в якій використовуються подвійний потенціал Морзе та потенціал Морзе між киснями; параметри моделі вибрано так, щоб пояснити різноманітні загальні влас\-тивості низки сполук з водневими зв'язками. З розрахунків у рамках цієї моделі випливає виникнення геометричного ефекту (ефекту Уббелоде): видовження водневого зв'язка при дейтеруванні, і це узгоджується з тим, що спостерігається в сегнетоелектриках з короткими водневими зв'язками. Використовуючи для параметрів потенціалів результати першопринципних розрахунків, розвинено одновимірну модель, в якій білінійні протон-протонні взаємодії розглядаються в наближенні середнього поля. Ця модель використовується для дослідження квантових ефектів у ядрах, які призводять до виникнення геометричного ефекту в кристалах KDP. Підхід ІТМК 
	дає змогу виявити, що протони тунелюють більш ефективно вздовж одновимірного ланцюжка, ніж дейтрони; це спричиняє появу сильного притягувального центра, який зменшує відстань між атомами киснів. Цей механізм, який ґрунтується на кореляції між тунелюванням і геометричними змінами водневих зв'язків, призводить до виникнення сильного геометричного ефекту в ланцюжку у впорядкованій фазі при низьких температурах, що добре узгоджується з експериментальними даними.
	\keywords сегнетоелектричний фазовий перехід, сегнетоелектрики з водневими зв'язками, метод інтегралів за траєкторіями у моделюванні Монте-Карло
	
\end{abstract}

 \lastpage
 \end{document}